\documentclass[epj,final]{svjour}

\usepackage{SIunits}
\usepackage{amsmath}
\usepackage{graphicx}
\usepackage{booktabs}
\usepackage[british]{babel}

\begin{document}

\title{A device for simultaneous spin analysis of ultracold neutrons}
\author{ 
  S.~Afach\inst{1,2,3}
  \and G.~Ban\inst{4}
  \and G.~Bison\inst{2}
  \and K.~Bodek\inst{5}
  \and Z.~Chowdhuri\inst{2}
  \and M.~Daum\inst{2}
  \and M.~Fertl\inst{1,2}\thanks{Now at University of Washington, Seattle WA, USA.}
  \and B.~Franke\inst{1,2}\thanks{Now at Max-Planck-Institute of Quantum Optics, Garching, Germany.}
  \and P.~Geltenbort\inst{6}
  \and Z.~D.~Gruji\'c\inst{7}
  \and L.~Hayen\inst{8}
  \and V.~H\'elaine\inst{2,4}\thanks{Now at LPSC, Grenoble, France.}\thanks{Corresponding author: helaine@lpsc.in2p3.fr}
  \and R.~Henneck\inst{2}
  \and M.~Kasprzak\inst{7}
  \and Y.~Kerma\"idic\inst{9}
  \and K.~Kirch\inst{1,2}
  \and S.~Komposch\inst{1,2}
  \and A.~Kozela\inst{10}
  \and J.~Krempel\inst{1}
  \and B.~Lauss\inst{2}
  \and T.~Lefort\inst{4}
  \and Y.~Lemi\`ere\inst{4}
  \and A.~Mtchedlishvili\inst{2}
  \and O.~Naviliat-Cuncic\inst{4}\thanks{Now at Michigan State University, East-Lansing, MI, USA.}
  \and F.~M.~Piegsa\inst{1}
  \and G.~Pignol\inst{9}
  \and P.~N.~Prashanth\inst{2,8}
  \and G.~Qu\'em\'ener\inst{4}
  \and M. Rawlik\inst{5}
  \and D.~Ries\inst{2,3}
  \and D.~Rebreyend\inst{9}
  \and S.~Roccia\inst{11}
  \and D.~Rozpedzik\inst{5}
  \and P.~Schmidt-Wellenburg\inst{2}
  \and N.~Severijns\inst{8}
  \and A.~Weis\inst{7}
  \and E.~Wursten\inst{8}
  \and G.~Wyszynski\inst{1,5}
  \and J.~Zejma\inst{5}
  \and G.~Zsigmond\inst{2}
}

\institute{
  ETH Z\"urich, Institute for Particle Physics, CH-8093 Z\"urich, Switzerland                                 %1
  \and Paul Scherrer Institute, CH--5232 Villigen-PSI, Switzerland                                       %2
   \and Hans Berger Department of Neurology, Jena University Hospital, D-07747 Jena, Germany                   %3
  \and LPC Caen ENSICAEN, Universit\'e de Caen, CNRS/IN2P3, F 14050 Caen, France                              %4
  \and Marian Smoluchowski Institute of Physics, Jagiellonian University, 30--059 Cracow, Poland              %5
  \and Institut Laue--Langevin, Grenoble, France                                                              %6
  \and Physics Department, University of Fribourg, CH--1700, Fribourg, Switzerland                            %7
\and Instituut voor Kernen Stralingsfysica, Katholieke~Universiteit~Leuven, B--3001 Leuven, Belgium      %8
  \and LPSC, Universit\'e  Grenoble Alpes, CNRS/IN2P3, Grenoble, France                                       %9
  \and Henryk Niedwodnicza\'nski Institute for Nuclear Physics, 31--342 Cracow, Poland                        %10
  \and CSNSM, Universit\'e Paris Sud, CNRS/IN2P3, Orsay campus, France                                        %11
}

\date{Received: date / Revised version: date}

\abstract{
  We report on the design and first tests of a device allowing for measurement of ultracold neutrons polarisation by means of the simultaneous analysis of the two spin components. The device was developed in the framework of the neutron electric dipole moment experiment at the Paul Scherrer Institute. Individual parts and the entire newly built system have been characterised with ultracold neutrons. The gain in statistical sensitivity obtained with the simultaneous spin analyser is $(18.2\pm6.1)\%$  relative to the former sequential analyser under nominal running conditions.
}

\maketitle

\section{Introduction}

Searches for the neutron electric dipole moment (nEDM) are challenging low energy experiments motivated by the potential discovery of new sources of CP violation \cite{Pospelov_2005,Dubbers_2011,Ramsey_Musolf_2013}. The nEDM experiment at the Paul Scherrer Institute (PSI) \cite{Baker2011} operates the RAL-Sussex-ILL spectrometer \cite{Baker_2014}, which has set the current most stringent nEDM limit \cite{Baker_2006}. The experiment is  connected to the new ultracold neutron source  \cite{Lauss_2012}. The spectrometer has been upgraded and since 2012 the PSI nEDM collaboration has been taking data.

The nEDM measurement is carried out using polarised ultracold neutrons (UCN) stored in a vessel and simultaneously exposed to an electric and a magnetic field. Ultracold neutrons are polarised by passing through a 5\,T superconducting magnet and then stored in the precession vessel where they precess freely. The difference between the Larmor frequencies measured with parallel ($\nu_{\parallel}$) and anti-parallel ($\nu_{\slash\!\!\!\parallel}$) static magnetic and electric fields gives the magnitude of the nEDM
\begin{equation}
  d_{\mathrm{n}}=\frac{h(\nu_{\slash\!\!\!\parallel}-\nu_{\parallel})}{4E}
\end{equation}
where $E$ is the electric field strength. The neutron Larmor frequencies are measured using Ramsey's technique of separated oscillating fields \cite{Ramsey_1950} which requires counting neutrons with spin up ($N^{\uparrow}$) and spin down ($N^{\downarrow}$) relative to the magnetic field direction, to produce the Ramsey interference pattern. The nEDM sensitivity derived from such a technique is given by
\begin{equation}
  \sigma_{d_{\mathrm{n}}}\simeq\frac{\hbar}{2\alpha TE\sqrt{N_{\mathrm{tot}}}}
  \label{eq:dn_statistic_error}
\end{equation}
where $T$ is the free precession time, $\alpha$ is the contrast of the Ramsey central fringe -- defined as the fractional amplitude of the spin-dependent signal modulation \cite{Golub1991}, and \mbox{$N_{\mathrm{tot}}=N^{\uparrow}+N^{\downarrow}$} is the total number of detected UCN. These last two parameters determine the figure of merit of the experiment and are directly related to the spin analysis and neutron detection system.

The spin analysis of the previous RAL-Sussex-ILL experiment was based on sequential counting of  $N^{\uparrow}$ and $N^{\downarrow}$ \cite{Baker_2014}, as described below. In order to improve the statistical sensitivity of the nEDM measurement and also to reduce possible spin dependent systematic effects, a new spin analyser allowing the simultaneous detection of spin up and spin down UCN has been developed. The three main objectives are: 1) to treat both spin components symmetrically, 2) to lower depolarisations and 3) to increase the number of detected UCN. This paper presents the simultaneous spin analysis device along with results from tests performed to evaluate its performance. Section \ref{sec:principle} describes the principle of the sequential and simultaneous spin analysis which provides motivation for the upgrade of the spin analyser. The detailed description of the analyser is given in Sec.~\ref{sec:Simultaneous spin analyser setup} and Sec.~\ref{sec:USSA subsystems characterisation} summarises its characterisation. Finally, the comparison of the sequential and the simultaneous analysers under typical measuring conditions for the nEDM experiment is presented in Sec.~\ref{sec:USSA test below the nEDM spectrometer}.

%%_________________________________________________________________________________________________%%
%% Motivations and principle %%

\section{Principles of spin analysis}
\label{sec:principle}

\begin{center}
  \begin{figure}[h]
    \centering
    \includegraphics[width=0.8\linewidth]{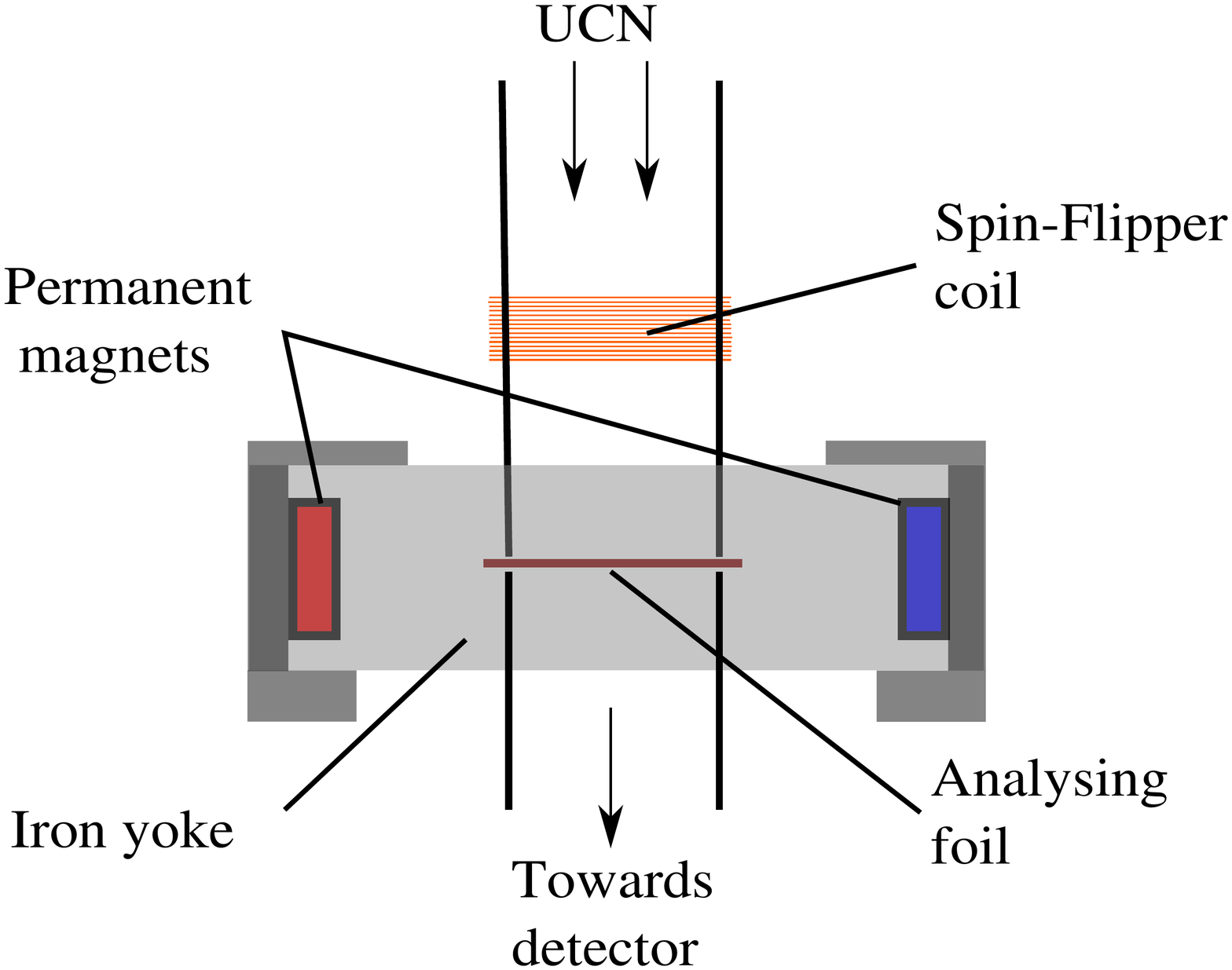}
    \caption{Scheme of the former spin analysing system. Ultracold neutrons fall down to the analysing foil which transmits only one spin state and then arrive at the detector. The spin component being transmitted can be changed with the spin-flipper.}
    \label{fig:spin_analysing_system}
  \end{figure}
\end{center}

A typical neutron spin analysis system consists of a magnetised analysing foil and an adiabatic spin-flipper (ASF) as shown in Fig.\,\ref{fig:spin_analysing_system},  where a scheme of the former sequential analyser is depicted.
The spin analysing foil is made of an iron layer magnetised up to saturation ($B_{\mathrm{sat}}\sim2$\,T) by the presence of a set of permanent magnets. Ultracold neutrons are reflected on the analysing foil if their kinetic energy is smaller than
\begin{equation}
U=V_F^{\mathrm{iron}}\pm|\mu_{\mathrm{n}}|B_{\mathrm{sat}}\simeq210\pm120\,\mathrm{neV}
\label{eq:foil_potential}
\end{equation}
where $V_F^{\mathrm{iron}}$ is the Fermi potential of iron, $\mu_{\mathrm{n}}$ the neutron magnetic moment and $B_{\mathrm{sat}}$ is the magnetic induction in the iron layer. The $+$ and $-$ signs correspond to spin up and spin down neutrons respectively. The potential energy $U$ produces a spin-dependent foil transmission,which we determine can yield a UCN spin analysing power of  95$\%$ for energies from 90\,neV to 330\,neV.

Since only one spin component can pass through the analysing foil, the ASF located upstream of the foil is used to flip the spin of neutrons in order to detect the spin up UCN. The principle of such an ASF is detailed in Refs.~\cite{Grigoriev_1997,Geltenbort_2009}.

\subsection{The sequential spin analysis}

In the sequential scheme of spin analysis, the detection of both spin components is performed by alternatively switching on and off the ASF according to a particular time sequence. The sequence is defined such that the same amounts of spin up and spin down UCN are detected from an initially unpolarised UCN sample. For example, in a typical nEDM measurement, UCN with spin down (ASF off) are counted during time intervals between 0 and 8\,s and between 33\,s and 50\,s after opening the precession chamber. The ASF is switched on 8\,s after opening the chamber so that the spin up component is analysed during the interval between 8\,s and 33\,s. The main drawback of this method is that one spin component is stored above the analysing foil while the other one is being counted. During this counting, the spin component which is stored undergoes depolarisations and losses in the rest of the nEDM apparatus, for instance in small gaps between guides. In addition, the sequential analysis can induce asymmetric losses and depolarisations of the two spin components and can therefore possibly create spurious systematic effects in the nEDM measurement. Although no such source of error has been identified yet, providing additional protection from spin-dependent systematic errors is clearly relevant to the next generation of neutron EDM experiments.

These considerations motivated the development of a simultaneous spin analysing system. Such a technique had been pioneered in early nEDM experiments at LNPI \cite{Altarev_1981} and pursued at PNPI \cite{Lasakov_2005}. However, the spin treatment with those systems was not symmetric in contrast to the method presented hereafter. Two prototypes of a symmetric two-arms system, having a different geometry than the device described here, had previously been studied, built and characterised  \cite{Rogel2009}.

\subsection{Principle of simultaneous spin analysis}

\begin{center}
  \begin{figure}[h]
    \centering
    \includegraphics[width=0.7\linewidth]{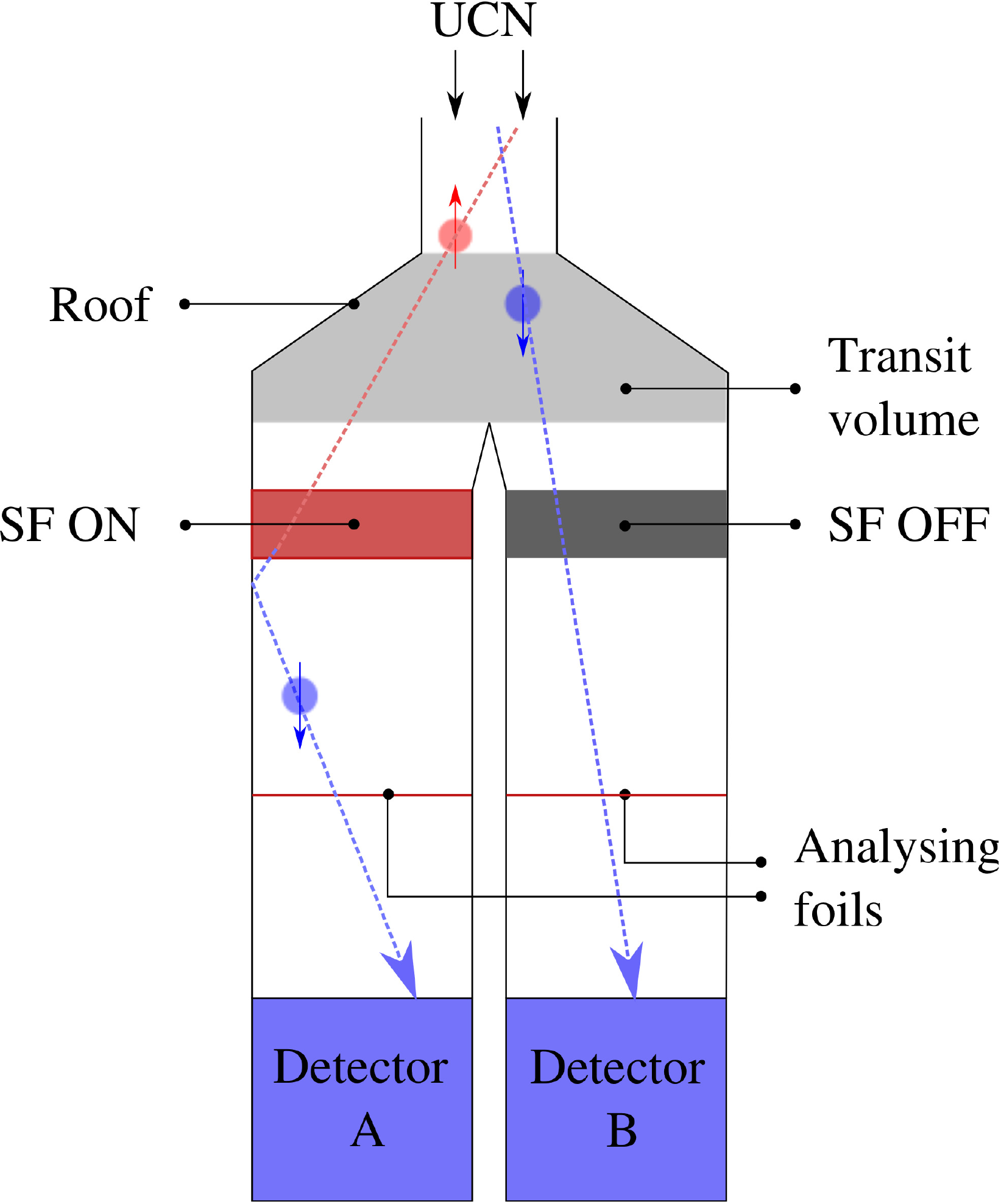}
    \caption{Scheme of the simultaneous spin analysing device. Each arm is made of a full spin analysis system including an ASF, a magnetised foil and a detector. The detector A (B) on the left (right) arm counts UCN which had initially their spin up (down) relative to the main magnetic field.}
    \label{fig:simultaneous_analysis_principle}
  \end{figure}
\end{center}

 The main idea for the simultaneous spin analysis is to use a complete spin analysis system as shown in Fig.\,\ref{fig:spin_analysing_system} for each spin component. This results in two identical arms, as shown in Fig.\,\ref{fig:simultaneous_analysis_principle}. In the left arm, where the ASF is on, the spin up component is analysed whereas in the right arm, where the ASF is off, the spin down component is measured. As a result, the storage time above the analysing foils is reduced and UCN losses as well as depolarisations are minimized. During operation, neutrons with initial spin down (up) will be reflected on the analysis foil of the left (right) arm. An important goal in the design of the simultaneous spin analyser  was to optimize the transport of ``wrong spin'' neutrons from one arm to the other. This is performed using a transit volume above the two arms with a particular geometry determined using \textsc{Geant4-UCN} simulations  \cite{G4UCN}.

%%_________________________________________________________________________________________________%%
%% Design using Geant4UCN simulations %%

\subsection{\textsc{Geant4-UCN} simulations}

\begin{center}
  \begin{figure}[h]
    \centering
    \includegraphics[width=0.7\linewidth]{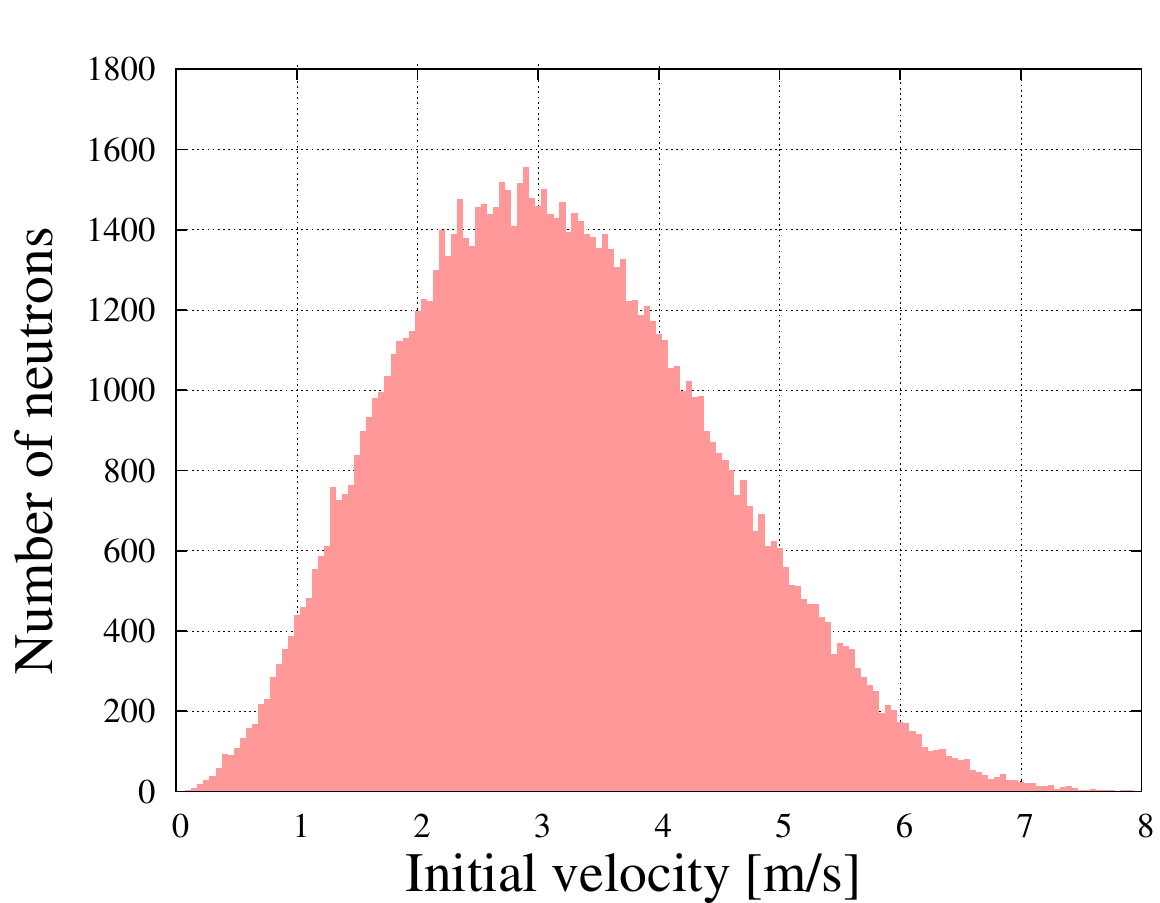}
    \caption{Initial velocity spectrum after  100\,s storage in the nEDM precession chamber used for simulations, taken from \cite{Rogel2009}. The corresponding maximal energy is 330\,neV.}
    \label{fig:simulation_velocity}
  \end{figure}
\end{center}

Simulations were performed mainly to assist in designing the simultaneous spin analysis system and also to compare relative performance of the sequential and the simultaneous systems \cite{Helaine2014}. The \textsc{Geant4-UCN} package, including UCN physics, was used. In each batch, a total of $10^5$ polarised or unpolarised UCN were uniformly generated in the precession chamber volume and were tracked until they were either lost or detected. The initial velocity spectrum is a simulated spectrum after 100\,s storage in the nEDM precession chamber taken from \cite{Rogel2009}, shown in Fig.\;\ref{fig:simulation_velocity}. The corresponding UCN energy distribution is defined in the range from 0 to 330\,neV with a maximum at about 40\,neV.

Both foils and depolarisations on  walls have been modelled by a $10^{-5}$ depolarisation per bounce probability. Ideal spin-flippers with a 100$\%$ spin-flip efficiency have been used.

The U-shaped Simultaneous Spin Analyser (USSA) was compared to the sequential analysing system considering both, the UCN detection efficiency and the spin asymmetry calculated from the detected spin up and spin down neutrons. The detection efficiency is defined as the ratio between the number of detected UCN and the initial number of generated UCN, assuming 100$\%$ efficient detectors. The asymmetry is defined as
\begin{equation}
\mathcal{A}=\left|\frac{N^{\uparrow}-N^{\downarrow}}{N^{\uparrow}+N^{\downarrow}}\right|
\end{equation}
and reflects the spin analysing power of the system. In the ideal case, the asymmetry is 100$\%$ for fully polarised UCN and 0$\%$ for an unpolarised UCN population.
The same two criteria have been used to optimize the geometry and the dimensions of the USSA. For example, the angle of the roof (45$^{\circ}$) and the angle of the quartz wedge (Fig.\;\ref{fig:USSA_mechanical_design}) between the two arms (30$^{\circ}$) were defined by maximising the detection efficiency.
The inner coating of the arms was chosen to be NiMo (85$\%$ nickel-15$\%$ molybdenum weight ratio) and the field in the analysing foils was set to 2\,T. Ideally, the vertical drop to the foils would be controlled to avoid analysing foil-crossings by the wrong spin component as well as neutron losses in the USSA. Practical constraints require a vertical drop of roughly 2m from the EDM cell to the analysing foil. Given the expected spectrum in USSA above the foils, it is clear that a higher Fermi potential than that of NiMo would be helpful in the USSA system to ensure high transport efficiency for neutrons reflected by the analysing foil, for example. We will return to this point in our evaluation of the USSA performance in Sec.'s \ref{sec:USSA subsystems characterisation} and \ref{sec:USSA test below the nEDM spectrometer}.

\begin{table}[h]
\caption{Detection efficiency, $\varepsilon_{\mathrm{det}}$, and asymmetry,  ${\mathcal{A}}$, for initially unpolarised and polarised UCN populations obtained from the simulations for the two schemes of spin analysis.
The quoted uncertainties are statistical.}
\label{tab:comparizon_analysers}
  \centering{\begin{tabular}{lcccc}
      \hline
      \hline
      Analysis & \multicolumn{2}{c}{Unpolarised} & \multicolumn{2}{c}{Polarised}\\
      scheme & ${\varepsilon_{\mathrm{det}}} [\%]$ & ${\mathcal{A}} [\%]$ & ${\varepsilon_{\mathrm{det}}} [\%]$ & ${\mathcal{A}} [\%]$ \\
      \hline
      Sequential &
      74.2(3) &  3.1(4)  & 74.2(3) & 96.9(1)\\
      Simultaneous &
      78.5(3) &  0.1(4)  &  78.8(3)  &  99.3(4)\\ 
      \hline
      \hline
  \end{tabular}}
\end{table}

The comparison between the performance of the USSA and the sequential analyser are summarised in Table~\ref{tab:comparizon_analysers}.
In the simulations, there was no slit included above the analysing foils. Therefore, the losses during storage of each spin component for the sequential analysis are lower than in reality. As a result, the $6\%$ increase of UCN detection efficiency using the USSA compared to the sequential analyser is likely underestimated. 

Those simulations have also been used to reveal mechanisms responsible for the analysing power decrease of the sequential system. First, a part of the UCN spin population transmitted by the foil, located in the volume between the ASF and the detector, is accounted for by the other spin population, just after the ASF is switched on or off, because of the UCN time of flight to go from the spin flip area and the detector. In addition, artificial spin-flips occur for the UCN population with the spin state not transmitted by the analyser: when those UCN are located between the ASF and the foil just before the ASF is switched on or off, they have crossed the ASF on, going towards the analysing foil -- with the associated spin-flip, and then they cross the ASF off after being reflected by the foil -- without spin-flip. As a result, the asymmetry obtained with the sequential analyser is reduced for polarised UCN. By tuning the time sequence, it is possible to compensate the two phenomena for an unpolarised UCN population. However, this uncomfortable configuration does not enable us to recover the full asymmetry for polarised UCN. In contrast, the symmetric treatment of both spin states in the USSA results in an asymmetry closer to unity and does not require any tuning.

Both the UCN detection efficiency and the asymmetry are improved in the symmetric treatment of spin states with the USSA as compared to the sequential analyser.

%%_________________________________________________________________________________________________%%
%% Simultaneous Spin Analyser building %%

\section{Simultaneous spin analyser setup}
\label{sec:Simultaneous spin analyser setup}

\begin{center}
  \begin{figure}[h]
    \centering
    \includegraphics[width=0.95\linewidth]{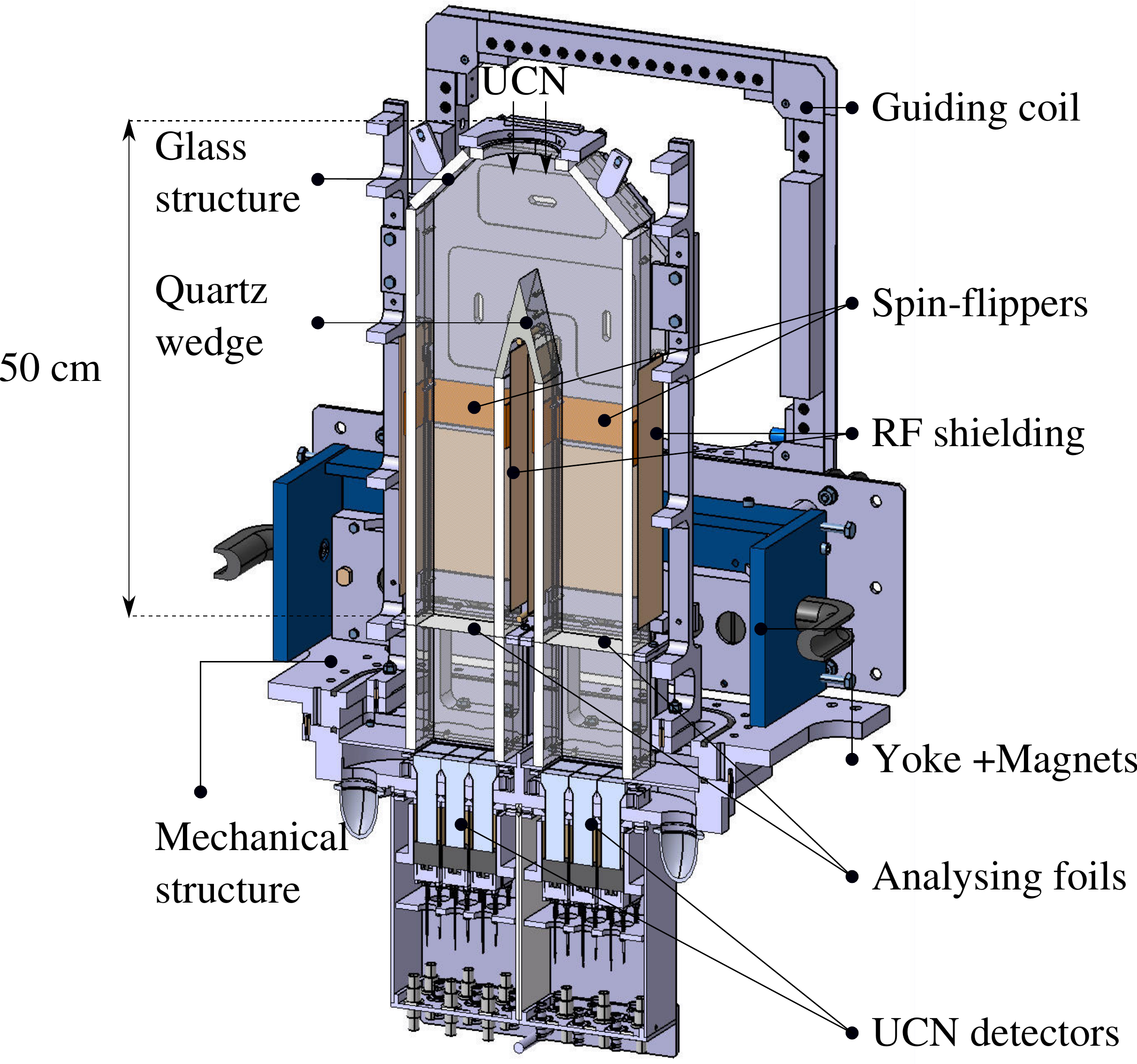}
    \caption{Vertical cut view of the USSA mechanical design. The aluminium vacuum chamber is omitted.}
    \label{fig:USSA_mechanical_design}
  \end{figure}
\end{center}

The mechanical design of the USSA is shown in Fig.\,\ref{fig:USSA_mechanical_design}. The full apparatus except the permanent magnets and the iron yoke, is located inside an aluminium vacuum chamber (not shown in Fig.\,\ref{fig:USSA_mechanical_design}). The permanent magnets and the soft iron yoke surround the chamber at the level of the analysing foils.

\subsection{UCN transport}

\begin{center}
  \begin{figure}[h]
    \centering
    \includegraphics[width=0.74\linewidth]{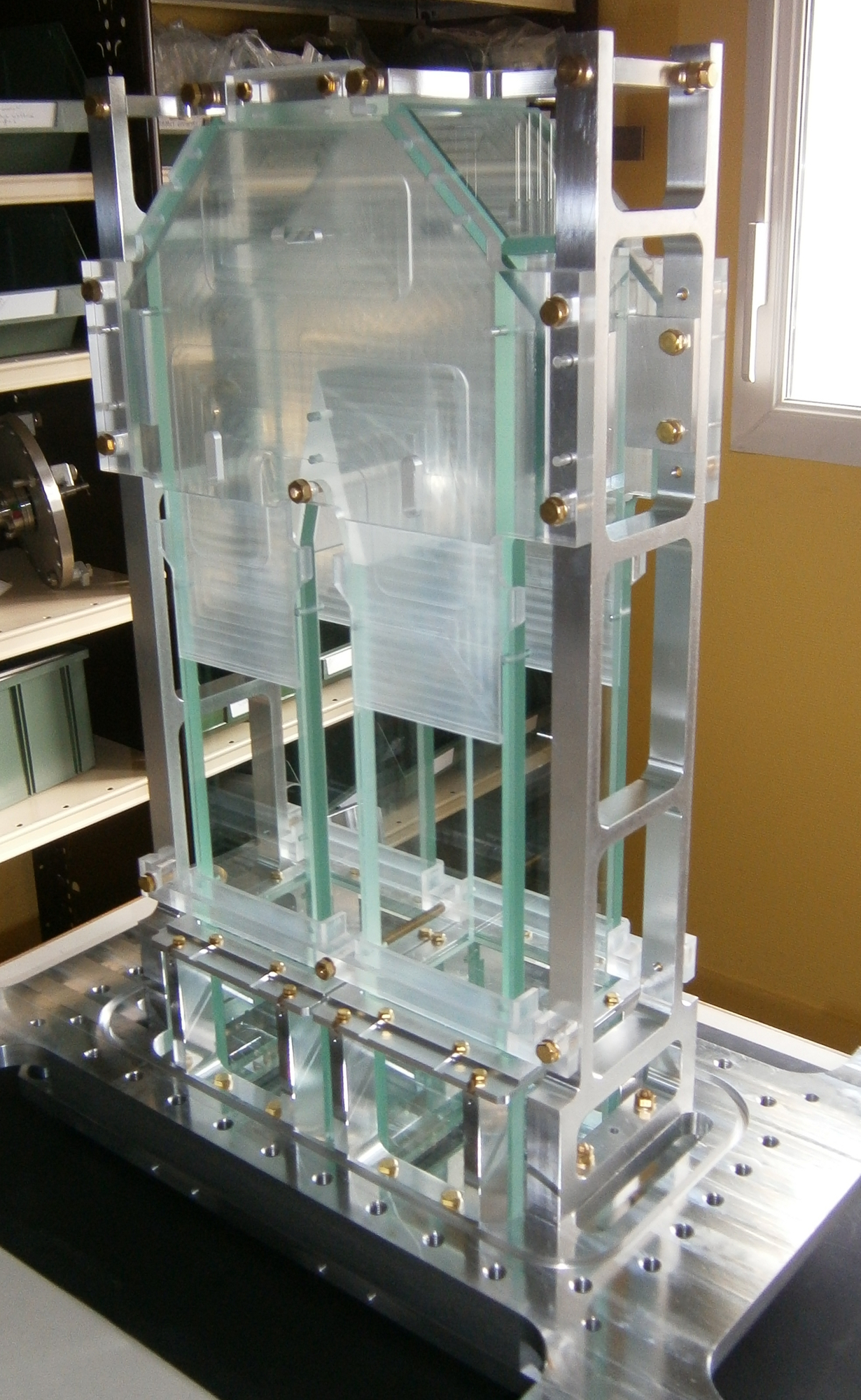}
    \caption{Bare float-glass structure held by the aluminium and PMMA support.}
    \label{fig:USSA_glass_structure}
  \end{figure}
\end{center}

Ultracold neutrons entering the USSA are guided toward the detectors using float glass walls. The small roughness of the glass material, of the order of a few nm, favours UCN transport along the arms. The glass structure (Fig.\,\ref{fig:USSA_glass_structure}) is coated with a thin layer ($300-500$\,nm) of sputtered NiMo which has a Fermi potential for neutrons of 220\,neV (for a 85/15, Ni/Mo weight ratio). The glass plates are assembled and tightened up by an external non-magnetic structure made of aluminium and PMMA\footnote{Polymethyl methacrylate}. The largest gaps between the glass plates, of about 100\,\micro m, are located at the top of the structure along a few cm, between the roof and the vertical walls. 

At the top, the roof makes an angle of 45$^{\circ}$ with respect to the arm axis. This angle  was obtained from the \textsc{Geant4-UCN} simulations, in order to favour UCN transport from one arm to the other. At the level of the analysing foils, the glass structure is split into two parts. This enables an easy exchange or removal of the analysing foils by moving down the lower part of the structure. Once mounted, a 200\,\micro m gap remains between the two glass parts. Considering both the 100\,\micro m gaps along USSA ridges and the 200\,\micro m gaps near the analysing foils, the total gap represents less than 0.15$\%$ of the total USSA surface. A wedge, with a 30$^{\circ}$ opening angle, is installed just below the entrance. In order to machine a sharp ridge on this piece, polished quartz has been used due to its hardness and its UCN guiding properties that closely match those  of glass. The wedge piece was also coated with NiMo.

\subsection{Spin transport and spin handling}

\subsubsection{Analysing foils and magnetisation system}

The analysing foils are made of a 400\,nm iron layer deposited on a 25\,\micro m aluminium foil. The magnetic field required to magnetise such layers is $\approx$ 50\,mT \cite{BVR12}. At saturation, the magnetic induction inside the iron layer is between 1.8\,T and 2\,T. This corresponds to up to a $\pm120$\,neV potential added to the Fermi potential of Fe depending on the UCN spin state, following Eq.~(\ref{eq:foil_potential}) \cite{Golub1991}.

\begin{center}
  \begin{figure}[h]
    \centering
    \includegraphics[width=0.9\linewidth]{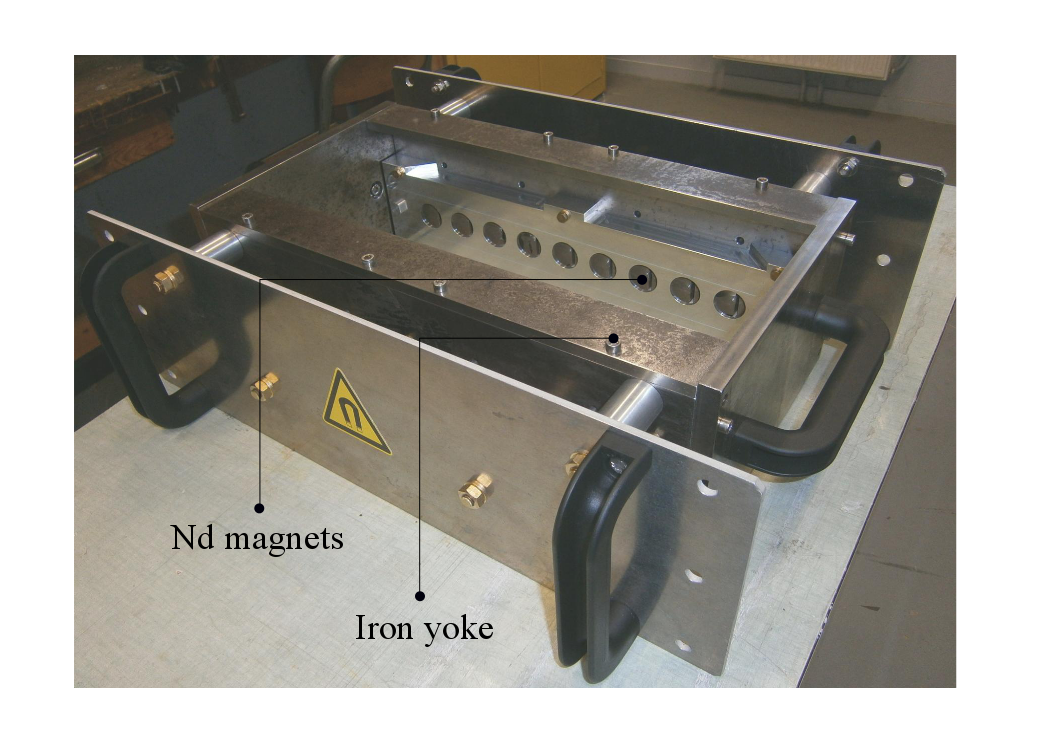}
    \caption{The magnetising system with the permanent magnets enclosed in the iron return yoke.}
    \label{fig:USSA_magnetisation_system}
  \end{figure}
\end{center}

The magnetising system (Fig.\,\ref{fig:USSA_magnetisation_system}) consists of a set of 40 Nd magnets (1.32\,T at the surface) enclosed in a rectangular return yoke made of soft iron. The return yoke is not symmetric with respect to the foil location in order to provide a suitable stray field to the spin-flippers.
The magnetising field at the position of the foils is in the $80-120$\,mT range, which ensures saturation of the magnetisation.

\subsubsection{Guiding fields}

\begin{center}
  \begin{figure}[h]
    \centering
    \includegraphics[width=0.95\linewidth]{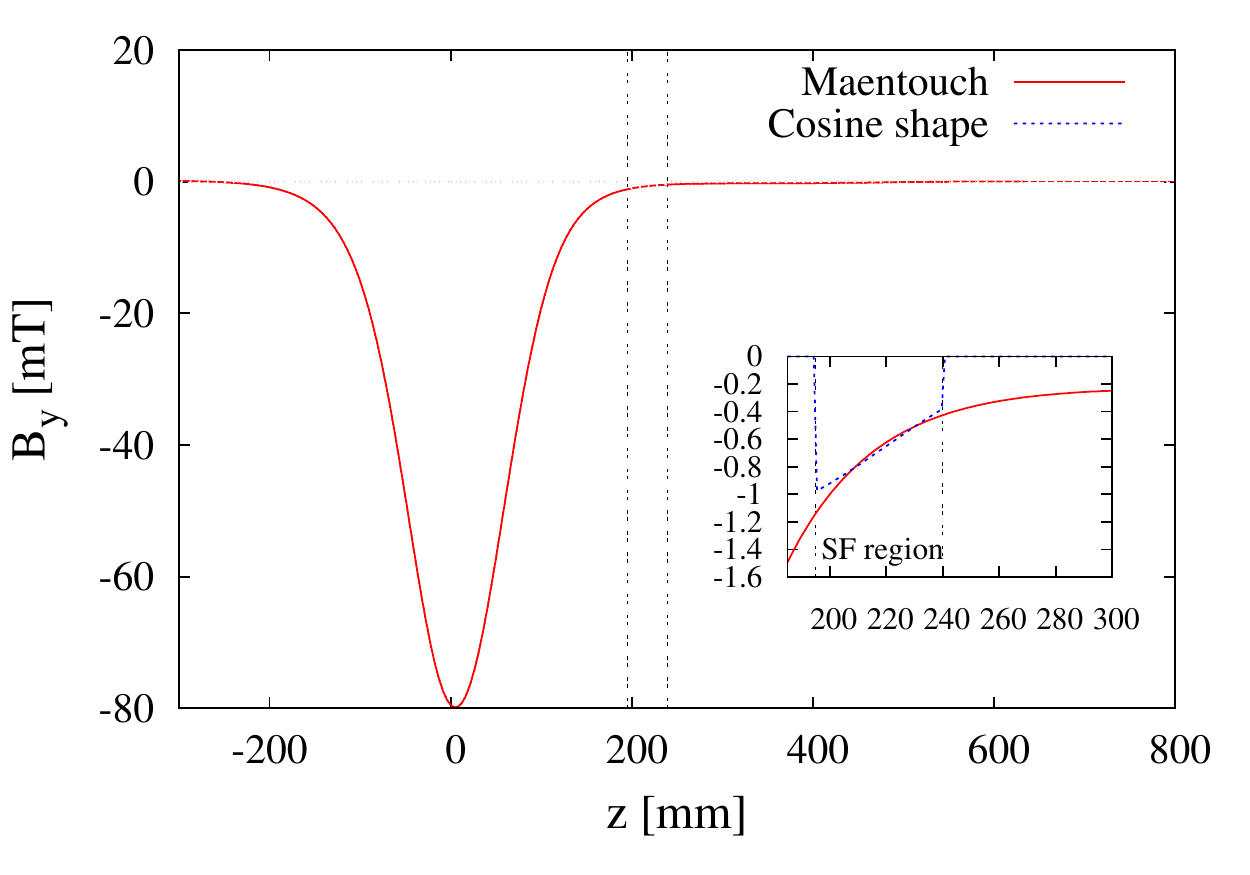}
    \caption{Red line: Calculation of the static magnetic field component perpendicular to the arm axis as a function of the vertical position. The field is created by the magnetisation system and the guiding coils. The analysing foils are located at $z=0$\,mm and the ASF is located between $z=195$\,mm and $z=240$\,mm. Blue dotted line: part of the cosine function modelling the static magnetic field in the sine-cosine model \cite{Grigoriev_1997} in the spin-flipper region. The insert shows a zoom near the ASF region.}
    \label{fig:magnetic_field_USSA}
  \end{figure}
\end{center}

From the entrance of the system down to the analysing foils, the UCN polarisation is maintained using the stray field of the magnetising system and a pair of saddle coils surrounding the upper part. The guiding field is perpendicular to the arm axis. The additional coils are required in order to suppress a region of otherwise zero field located off-axis, above the ASFs. The strength of the additional transverse guiding field is of the order of 400\,\micro T at the middle of the upper part. The field created by the magnetising system and the guiding coils is shown in Fig.\,\ref{fig:magnetic_field_USSA} along the arm axis. It was calculated using the Maentouch code \footnote{Maentouch is a custom boundary element code to model 3D magnetic fields developed by G. Qu\'em\'ener (CNRS).}. The result was confirmed by field map measurements. 

\subsubsection{Adiabatic spin-flippers}

Each ASF contains one coil producing an axial RF field. The transverse field gradient is created by the magnetising system and by two additional coils. The system was designed  to adiabatically flip the UCN spins.

\begin{center}
  \begin{figure}[h]
    \centering
    \includegraphics[width=0.8\linewidth]{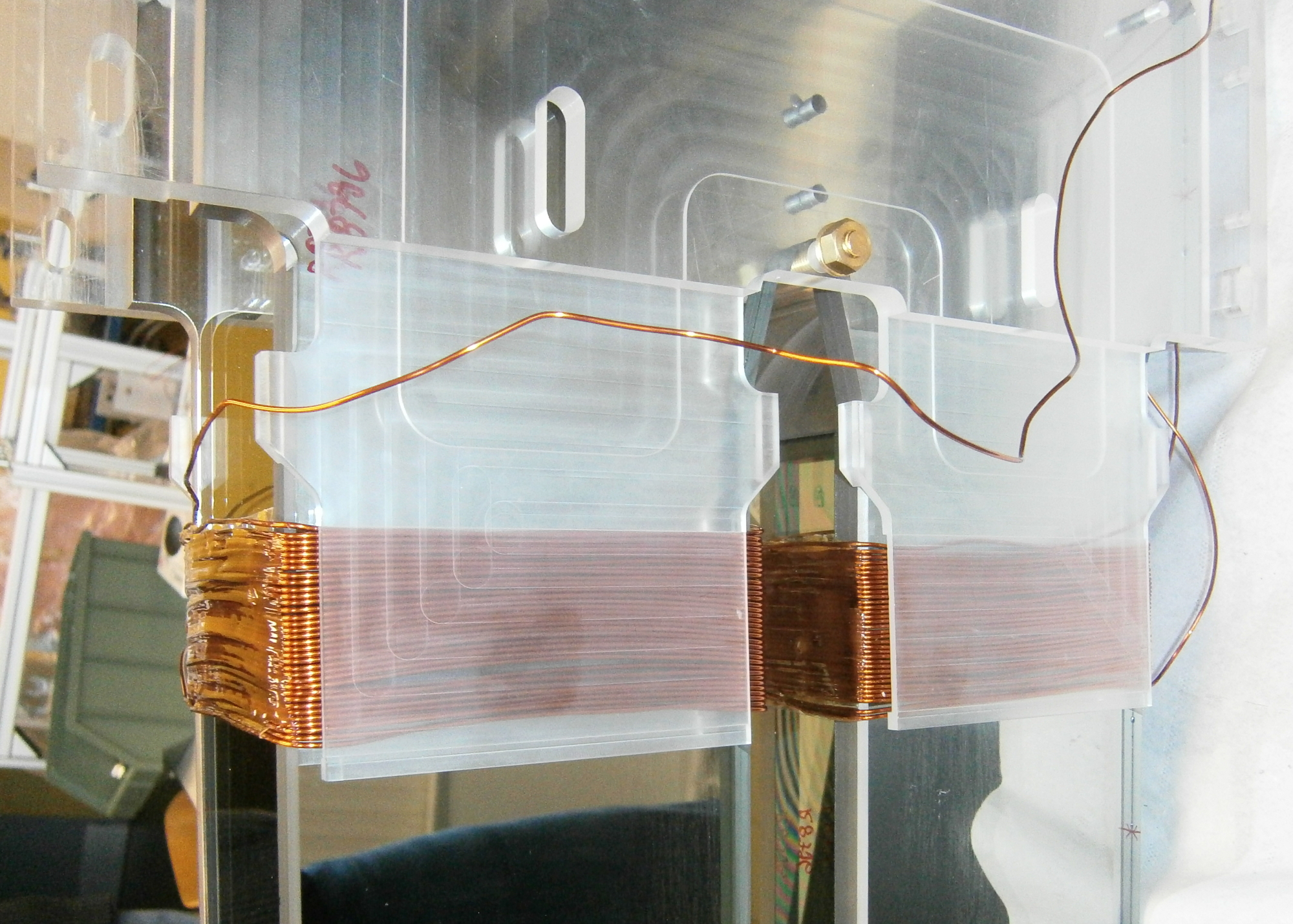}
    \caption{Spin-flipper coils, wound around the two arms. The copper RF shieldings (removed for clarity) are located 3\,mm around the spin-flippers.}
    \label{fig:USSA_spin-flippers}
  \end{figure}
\end{center}

\begin{center}
  \begin{figure}[h]
    \centering
    \includegraphics[width=0.98\linewidth]{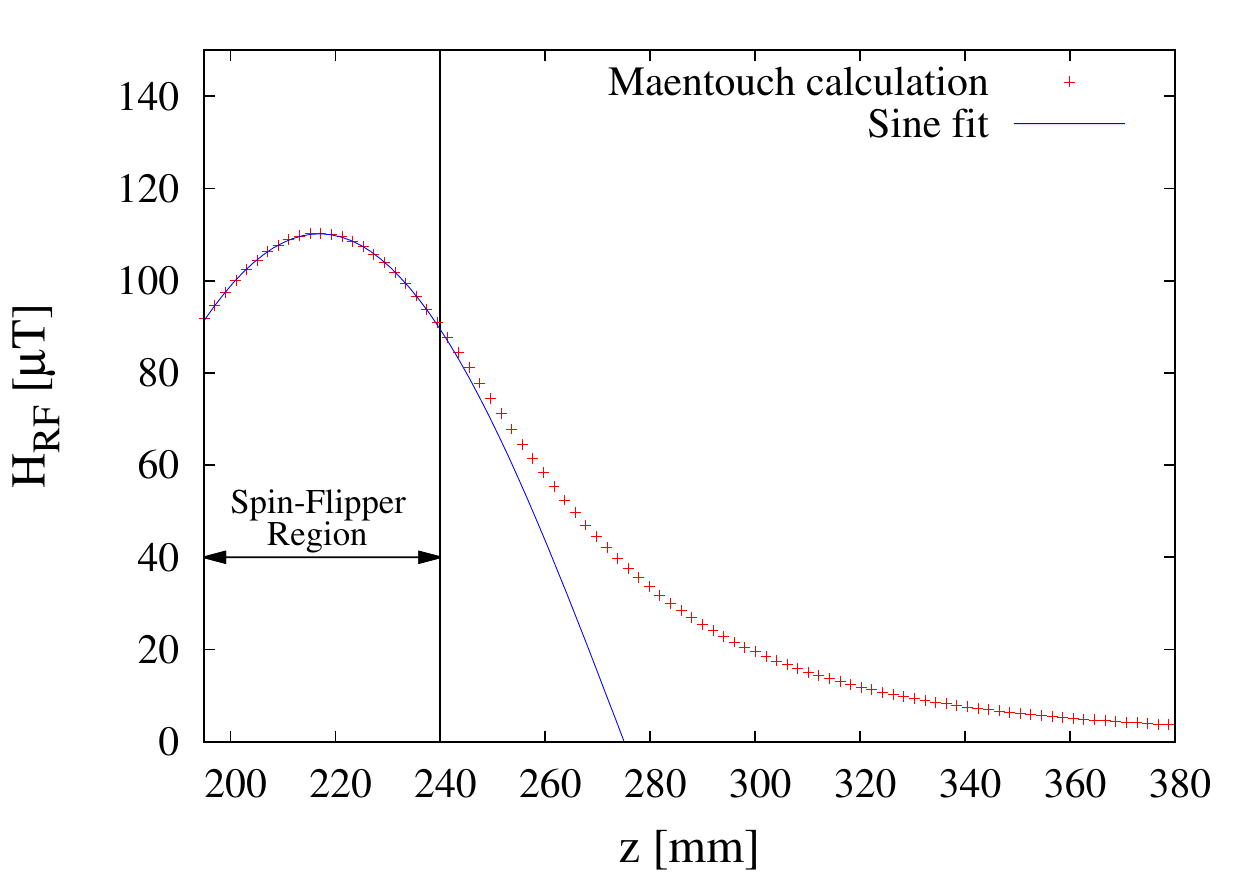}
    \caption{Adiabatic spin-flipper radio-frequency magnetic field amplitude along the arm axis. Red crosses: analytic calculation. Blue line: sinusoidal fit in the region of the spin flipper.}
    \label{fig:SF_field_shape}
  \end{figure}
\end{center}

The ASF have a square shape of 10\,cm side and a height of 4.4\,cm (Fig.\,\ref{fig:USSA_spin-flippers}). 
The frequencies of the ASF RF fields were set to 20\,kHz in order to fulfil the resonance condition at the ASF centre where the static field is 0.7\,mT.
The variation of the static magnetic field in the region of the spin-flipper follows approximately a cosine function (Fig.\,\ref{fig:magnetic_field_USSA}) and the variation of the RF transverse magnetic field amplitude approximately follows a sine function (Fig.\,\ref{fig:SF_field_shape}). These are the two conditions for the application of the sine-cosine model described in Ref. \cite{Grigoriev_1997}. Within this model the spin-flip probability $p_{\mathrm{flip}}$ as a function of the adiabaticity parameter, $k$, is given by
\begin{equation}
p_{\mathrm{flip}}=1-\frac{\sin^2{\left(\pi\sqrt{1+k^2}/2\right)}}{1+k^2}~\,.
\label{eq:sf_probability}
\end{equation}
The adiabaticity parameter quantifies the ability of the UCN spin to follow the magnetic field. It is defined as the ratio between the neutron Larmor frequency and the relative change of the magnetic field
\begin{equation}
k=\frac{\gamma_{\mathrm{n}}H_{\mathrm{RF}}^2}{v_{\mathrm{n}}\partial B_{\mathrm{y}}/\partial z}
\end{equation}
where $H_{\mathrm{RF}}$ is the RF field amplitude and $v_{\mathrm{n}}$ is the neutron velocity. If the field change is much smaller than the Larmor frequency ($k\gg1$), then the neutron spin is able to adiabatically follow the magnetic field.
The effective amplitude of the RF field at the ASF centre is $H_{\mathrm{RF}}=115$\,\micro T: the amplitude contributing to the adiabatic spin-flip is half of the linear oscillating field amplitude \cite{Abragam_1961}. Because of the height difference between the precession chamber and the analysing foils, the UCN speed is boosted up to about 8\,m/s. With a transverse field gradient of 0.15\,mT/cm, the adiabaticity parameter is $k\simeq20$. Such an adiabaticity coefficient leads to a spin-flip probability close to 100$\%$ according to Eq.\,(\ref{eq:sf_probability}). It has to be noticed that the sine-cosine model does not take into account all depolarisation mechanisms, such as depolarisations due to wall collisions in the presence of field gradients \cite{Gamblin_1965,Steyerl2012}, contributing to a decrease of the spin-flipper efficiency.

\subsubsection{RF shielding}

The simultaneous spin analyser is operated with one ASF on and the other off. Any RF cross-talk between the two arms has to be cancelled. The residual RF field in the non-active arm (ASF off) is suppressed using a 1\,mm thick copper shield placed around each arm. With such a shielding, the maximal RF field amplitude is reduced from 7.5\,\micro T to less than 0.3\,\micro T at the closest position. As a result, the spin-flip probability in the non-active arm decreased from 1.9$\%$ to a negligible level  sub-ppm, for a minimum UCN speed of 5\,m/s.

\subsection{UCN detection}

At the bottom of the arms, UCN are detected with two detector arrays based on $^6$Li glass scintillators \cite{Ban_2009,Lefort_2016}. Such detectors are able to handle counting rates up to a few $10^6$ counts/s and have comparable detection efficiency as a standard $^3$He gas detector. Each signal of the array is read out using a customised digital data acquisition system \cite{Faster}.

%%_________________________________________________________________________________________________%%
%% Characterisation of the USSA subsystems %%

\section{Characterisation of USSA subsystems}
\label{sec:USSA subsystems characterisation}

The performances of the spin-flippers, the analysing foils, as well as the transmission were measured at the West-2 beam port of the PSI UCN source \cite{Lauss_2012}. These tests were carried out with the UCN source operating in a mode delivering 3\,s proton beam pulses every 360\,s.

\subsection{Experimental setup}

\begin{center}
  \begin{figure}[h]
    \centering
    \includegraphics[width=0.9\linewidth]{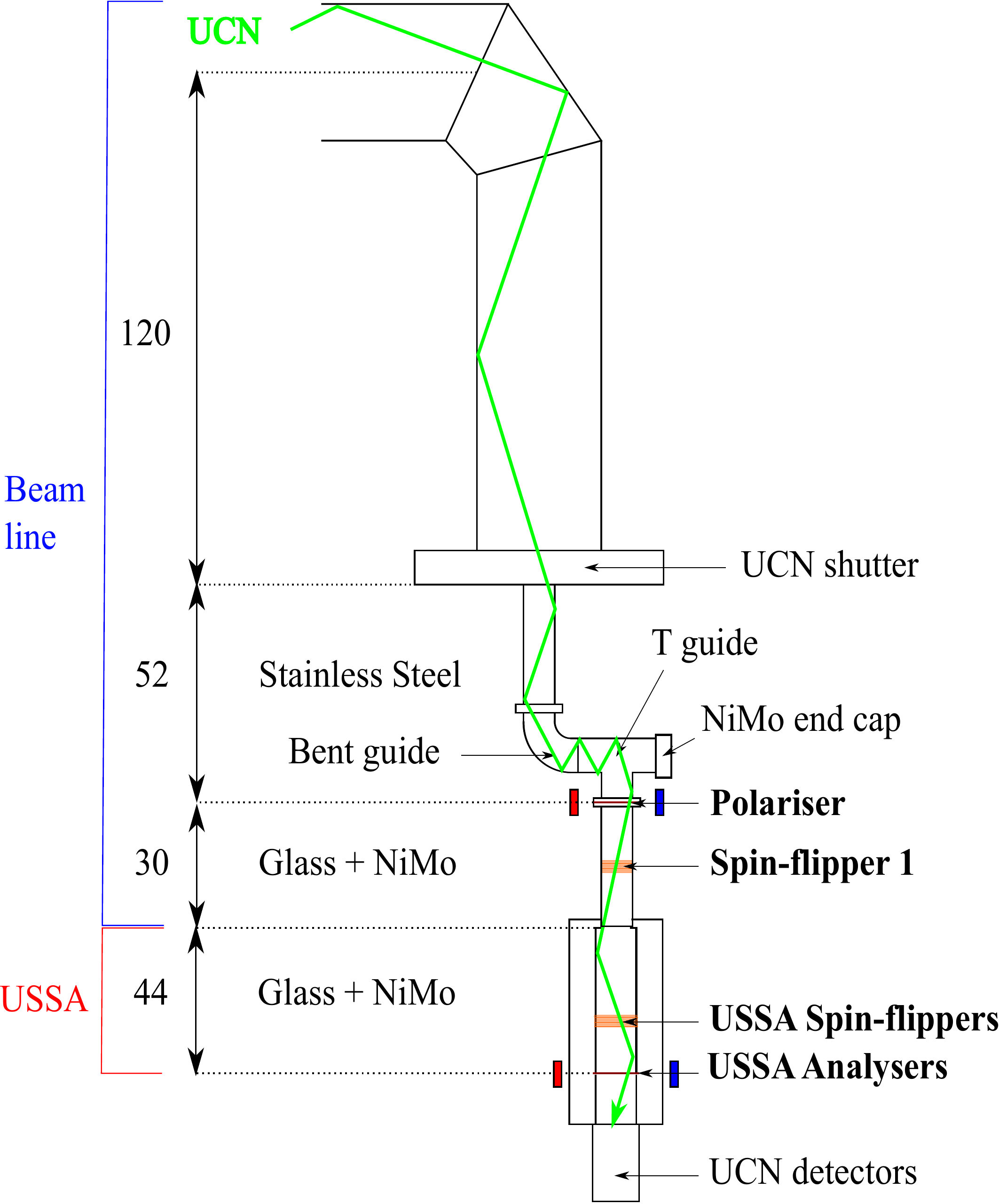}
    \caption{Measurement setup at the West-2 beam line used to characterise the USSA subsystems. Dimensions are in cm.}
    \label{fig:West2_beam_line}
  \end{figure}
\end{center}

The beam line setup is shown in Fig.\,\ref{fig:West2_beam_line}. Ultracold neutrons are extracted from the top of the UCN storage vessel, then follow the bend and fall down into the USSA. A stainless steel T-shaped guide was used in order to suppress the higher energy component from the UCN spectrum. A magnetised iron layer on an aluminium substrate was mounted immediately after the T-shaped guide in order to perform tests with polarised UCN. A first spin-flipper (noted spin-flipper 1) downstream of the polariser was installed to select the spin component of interest.

The number of detected UCN was normalised relative to the number of UCN simultaneously detected on the West-1 beam line to take into account the UCN source fluctuations.

The UCN energy spectrum range was estimated at the level of the analysing foils. The lowest energy, calculated from the height difference between the West-2 beam port and the USSA entrance is of about 230\,neV. A separate foil transmission measurement of a NiMo coated foil showed that 10$\%$ of the detected UCN had a kinetic energy higher than 330\,neV at the foil level. This represents a significantly larger UCN fraction with energies greater than the analysing potential for the iron foil than what is expected for the nEDM cell after a usual 200\,s storage time (about 0.2\% above the 330 neV foil level). We therefore expect the measured properties of the subsystem components such as the transmission and the reflection probability for ``wrong-spin'' UCN to be reflected by the foil to be greater for the actual nEDM experiment, and treat the quantities in Sec. \ref{sec:USSA subsystems characterisation} as a baseline measurement
to establish representative performance values.

\subsection{Tests with unpolarised UCN}

The USSA device was first characterised with unpolarised UCN. The set of measurements reported in this section was therefore carried out without the polariser.

\subsubsection{USSA transmission}

Transmission measurements were performed using a single UCN detector and the USSA. Both systems were installed at the same location and the counting rates were measured and compared. The USSA analysing foils were removed. The transmission of each arm was found to be $(83.2\pm0.7)\%$. The same measurement performed with USSA located 30\,cm lower gave a transmission of $(80.8\pm0.6)\%$. This indicates that the USSA transmission can be further increased using a coating with a higher Fermi potential. This conclusion is especially true with the analysing foils in place, because of the short storage time required in the volume above the foils.

\subsubsection{USSA detection asymmetry}
\label{subsubsec:USSA detection asymmetry}

The USSA detection asymmetry arises from the combination of the different arms transmission and the different detectors efficiency. Tests performed at the Institut Laue-Langevin (ILL) yielded a $(0.5\pm1.5)\%$  relative difference between the two arms transmission and a $(1.1\pm0.4)\%$ relative difference of the two UCN detectors efficiencies.

\begin{center}
  \begin{figure}[h]
    \centering
    \includegraphics[width=0.98\linewidth]{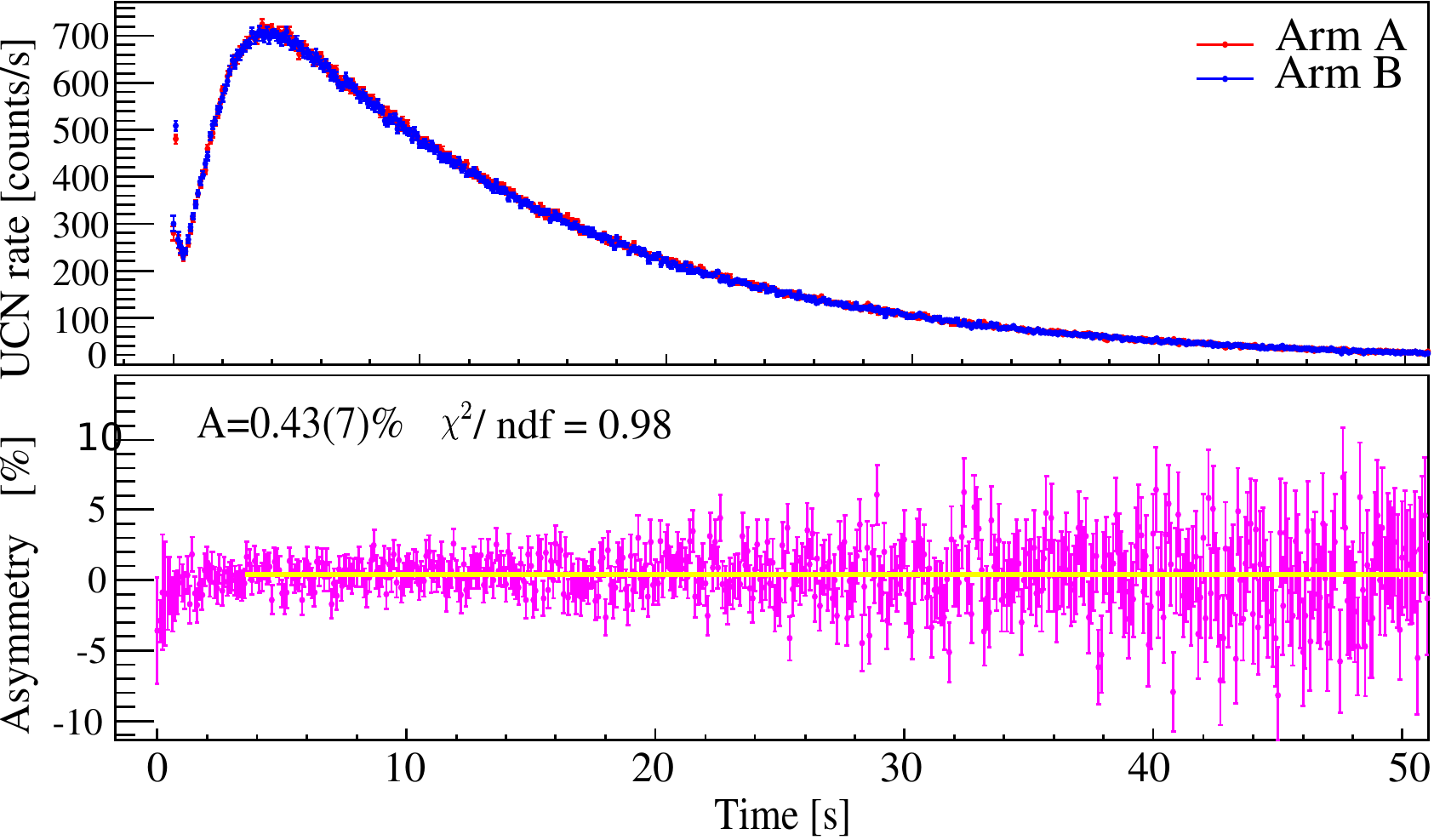}
    \caption{UCN counting rate (upper panel) and detection asymmetry (lower panel) as a function of time. The reference time is given by the timing of the primary proton beam kick. The measurement was performed without polarising and analysing foils.}
    \label{fig:west2_USSA_symmetry}
  \end{figure}
\end{center}

The raw UCN rates ($n_A$ and $n_B$) measured in arms A and B and the resulting asymmetry $\left(n_A-n_B\right)/\left(n_A+n_B\right)$ are shown in Fig.\,\ref{fig:west2_USSA_symmetry}. The few percent asymmetry observed during the first 2\,s is due to the remaining very cold neutrons coming from the UCN source. They were not included in the calculation of the mean detection asymmetry estimated to $(0.43\pm0.07)\%$.
The same measurement was performed with the two spin analysers installed in the USSA in order to study the possible influence of the analysing foils. Both ASF were off. The measured asymmetry was unchanged and amounted to $(0.40 \pm 0.11)\%$. This means that possible structural asymmetries between the two USSA arms do not impact the measured asymmetries at a level larger than $0.5\%$.

\subsubsection{Fraction of UCN detected after reflection on the opposite analyser}
\label{subsubsec:Proportion of detected UCN after reflection on the other analyser}

\begin{center}
  \begin{figure}[h]
    \centering
    \includegraphics[width=0.8\linewidth]{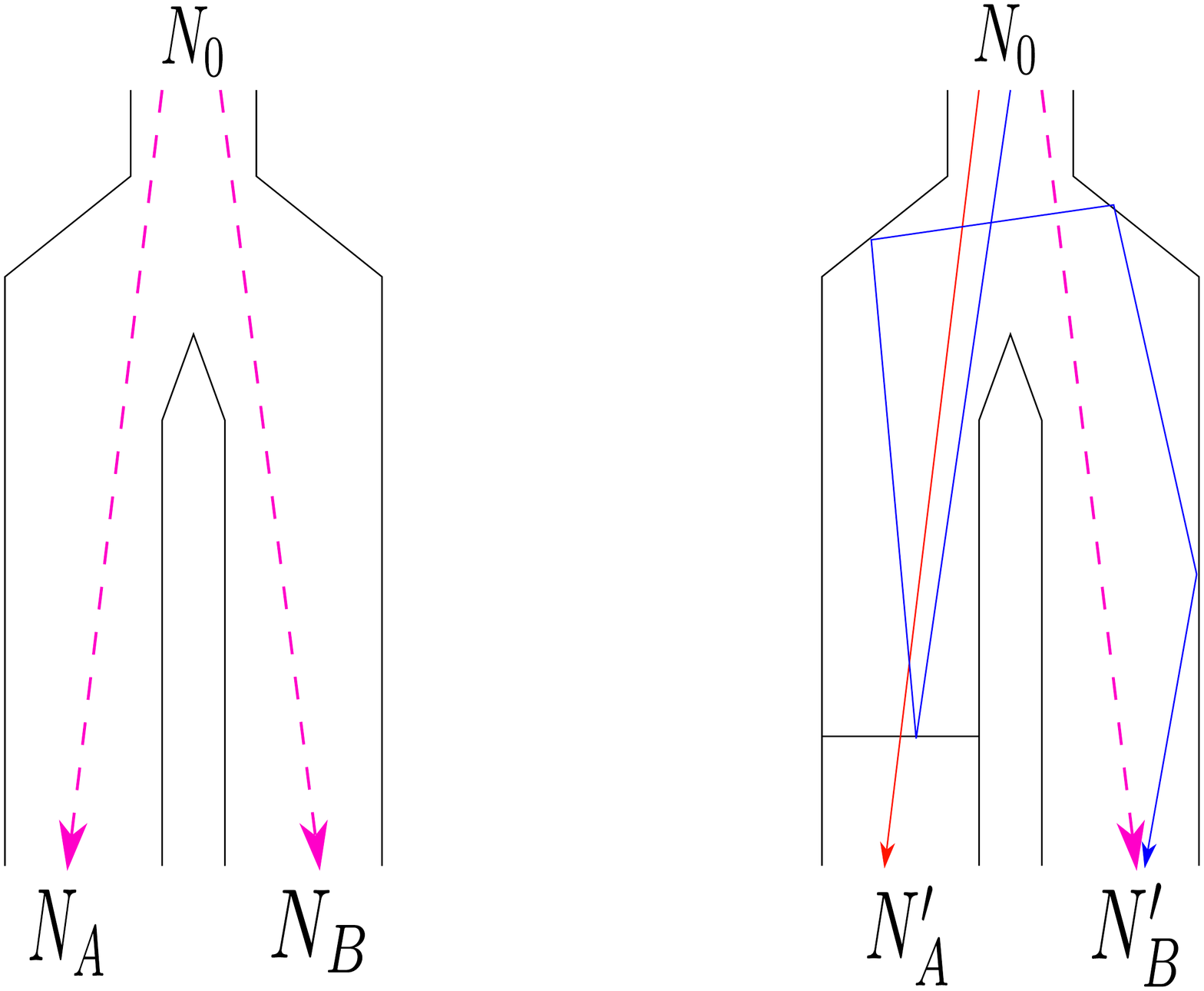}
    \caption{Configurations used to measure the UCN reflection probability from one arm to the other before being detected. $N_A$, $N_A'$, $N_B$ and $N_B'$ are the number of detected UCN in the two configurations. The UCN entering the device at the top are unpolarised. Blue and red lines represent spin down and spin up UCN respectively, while dashed purple represents both spin up and spin down UCN. On the left panel, there is no analysing foil while there is a single foil in the USSA on the right panel.}
    \label{fig:USSA_reflections_configuration_West2}
  \end{figure}
\end{center}

During operation, on average, half of each UCN spin population enters directly the arm dedicated to its analysis. The other half enters the other arm where it is expected to be reflected by the analysing foil. This section describes the measurement of the probability that a UCN going into the ``wrong arm'' bounces back to the correct arm where it is finally detected. In order to perform such a measurement, the two configurations shown in Fig.\,\ref{fig:USSA_reflections_configuration_West2} have to be considered.
Without analysing foil, the number of UCN detected in each arm are identical (Sec.~\ref{subsubsec:USSA detection asymmetry})
\begin{equation}
  N_A\simeq N_B=p_{T} N_0/2
  \label{eq:West2_conf1}
\end{equation}
where $N_0$ is the number of UCN entering the device at the top and $p_{T} $ is the transmission efficiency. With a spin analysing foil installed in one arm, the number of detected UCN, assuming a perfect analysing power of the foils, is given by
\begin{equation}
  \begin{cases}
    N'_A = p_{T}  N_0 \left(1 - p_{\mathrm{abs}}\right)/4\\
    N'_B = p_{T}  N_0 \left(2 + p_{\mathrm{ref}}\right)/4
  \end{cases}
  \label{eq:West2_conf2}
\end{equation}
where $p_{\mathrm{ref}}$ is the probability that a UCN is reflected on the analysing foil and is subsequently detected in the other arm, and $p_{\mathrm{abs}}$ is the absorption probability in the analysing foil. Using Eqs.\,(\ref{eq:West2_conf1}) and (\ref{eq:West2_conf2}), these probabilities are
\begin{equation}
  \begin{cases}
    p_{\mathrm{ref}} = 2\left( {N_B'}/{N_B}-1 \right)=(33.6\pm3.1)\%\\
    p_{\mathrm{abs}} = 1-{2N_A'}/{N_A} = (2.7\pm1.3)\%
  \end{cases}
  \label{eq:pref}
\end{equation}
The measured absorption probability in the analysing foil is in the same range as the one calculated from the absorption cross section for an Al foil of 25\,\micro m thickness ($1-3\%$ in the UCN energy range).

\subsection{Tests with polarised UCN}

For the simultaneous spin analysis, UCN were polarised through a magnetised foil installed upstream of the USSA as shown in Fig.\,\ref{fig:West2_beam_line}. The polarisation was measured using the method described in Ref.~\cite{Serebrov_1995}, developed for cold neutron beam experiments. This method is therefore well suited for single pass measurements performed in this section. The polarisation efficiency in such a configuration was previously measured to be $P_0=(70.6\pm3.2)\%$ using a dedicated beam line.

\subsubsection{Spin-flippers efficiencies}

The ASF efficiency $f$ of each arm has been measured using only a single foil in the arm of interest. Such a configuration allows independent measurements to be carried out avoiding multiple reflections between arms. Using the transmission matrix formalism  \cite{Serebrov_1995}, the USSA ASF efficiencies can be written as
\begin{equation}
f_{\mathrm{A}/\mathrm{B}}=\frac{N_{11}-N_{01}}{N_{00}-N_{10}} \text{ ,}
\end{equation}
where $N_{ij}$ is the number of UCN detected for an ASF configuration defined by indices $i$ and $j$. These indices correspond respectively to the state of the ASF 1, located upstream of the USSA, and the state of the ASF studied in the USSA. An index 0 corresponds to an ASF off and an index 1 to an ASF on. Four different configurations were set up from which the ASF efficiency of each arm was estimated: $f_{\mathrm{A}}=(97.0\pm1.2)\%$ and $f_{\mathrm{B}}=(97.1\pm0.9)\%$. These efficiencies do not change when the position of the device is lowered by 30\,cm. The $2-3\%$ difference relative to the maximal efficiencies can be due to either a non-adiabatic region in the ASF or to multiple reflections on the analysing foil and walls of the arm of interest inducing additional depolarisations.

\subsubsection{Spin-flip cross-talk}

Radio-frequency cross-talk measurements were carried out using an analysing foil in the non-active arm (ASF off) and by changing the ASF state of the opposite arm. The number of detected UCN in the non-active arm was then recorded. The relative variation in the count rate between the two ASF states was $(0.15\pm0.62)\%$. In conclusion, no cross-talk was observed between the USSA arms at this level of precision.

\subsubsection{USSA analysing power}
The analysing power of the foils was first separately measured with the configurations used for the ASF efficiency measurements and then the USSA spin analysing power was estimated. The analysing power is defined as the asymmetry response for an idealised 100$\%$ polarised beam. The product between the initial polarisation $P_0$ and the foil analysing power $P_{\mathrm{A}/\mathrm{B}}$ is given by \cite{Serebrov_1995}
\begin{equation}
  P_0\times P_{\mathrm{A}/\mathrm{B}}=\frac{N_{00}-N_{10}}{f_{\mathrm{A}/\mathrm{B}}N_{00}+N_{01}} \text{.}
\end{equation}
Both foils had the same analysing power at the 4$\%$ level: $P_{\mathrm{A}}=(91.0\pm4.4)\%$ and $P_{\mathrm{B}}=(89.7\pm4.3)\%$. Therefore, the spin is treated in a symmetric way by the USSA analysers.

For the second measurement, the two analysing foils were installed in the USSA. One ASF was on and the other was off, as it is supposed to operate for the nEDM experiment. The formalism presented in \cite{Serebrov_1995} holding for a single-pass geometry, the analysing power measured in this configuration was therefore an effective one. This analysing power was measured to be $(83.5\pm4.0)\%$. It was smaller than  with a single foil since bounces between arms lead to multiple foil reflections and ASF crossings.

%%_________________________________________________________________________________________________%%
%% USSA test in nEDM conditions %%

\section{USSA test below the nEDM spectrometer}
\label{sec:USSA test below the nEDM spectrometer}

\begin{center}
  \begin{figure}[h]
    \centering
    \includegraphics[width=0.8\linewidth]{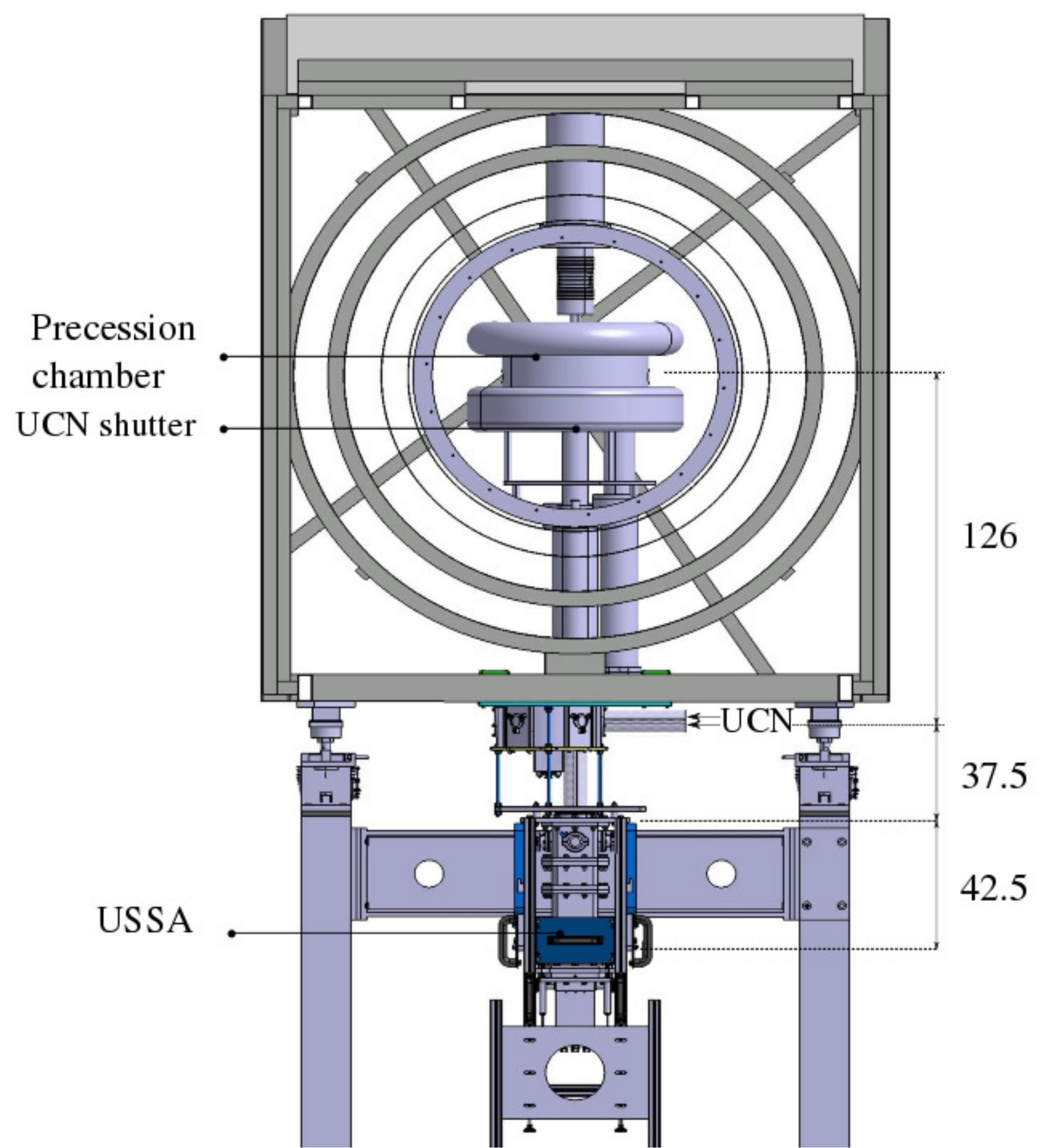}
    \caption{Simultaneous spin analyser integration in the nEDM apparatus. Ultracold neutrons come from the source horizontally and are then guided towards the precession chamber. After storage, the shutter is open and UCN fall towards the USSA. Dimensions are in cm.}
    \label{fig:USSA_integration_nEDM}
  \end{figure}
\end{center}

Finally, the USSA was installed below the nEDM spectrometer as shown in Fig.\;\ref{fig:USSA_integration_nEDM}. The USSA height allows all UCN to be analysed by the foils: the 2\,m vertical drop between the storage vessel and the foils ensures to be above the 90\,neV lower bound given by Eq.\;\eqref{eq:foil_potential} and the UCN spectrum softening in the nEDM storage vessel ensures UCN energies to be lower than the 330\,neV upper bound.

The reflection probability $p_{\mathrm{ref}}$ was first estimated with the method presented in Sec.~\ref{subsubsec:Proportion of detected UCN after reflection on the other analyser}. The precession chamber was filled during 35\,s with polarised UCN which were stored for 50\,s. Then, the shutter was opened and UCN fell down into the USSA located 1.6\,m below. The measured reflection probability was $(52.8\pm2.8)\%$. Hence, about half of UCN going into the arm dedicated to the opposite spin state detection were finally detected in their dedicated arm. This reflection probability is significantly higher than measured in our sub-system characterisation (Eq.\;(\ref{eq:pref})), as expected.

\begin{center}
  \begin{figure}[h]
    \centering
    \includegraphics[width=0.98\linewidth]{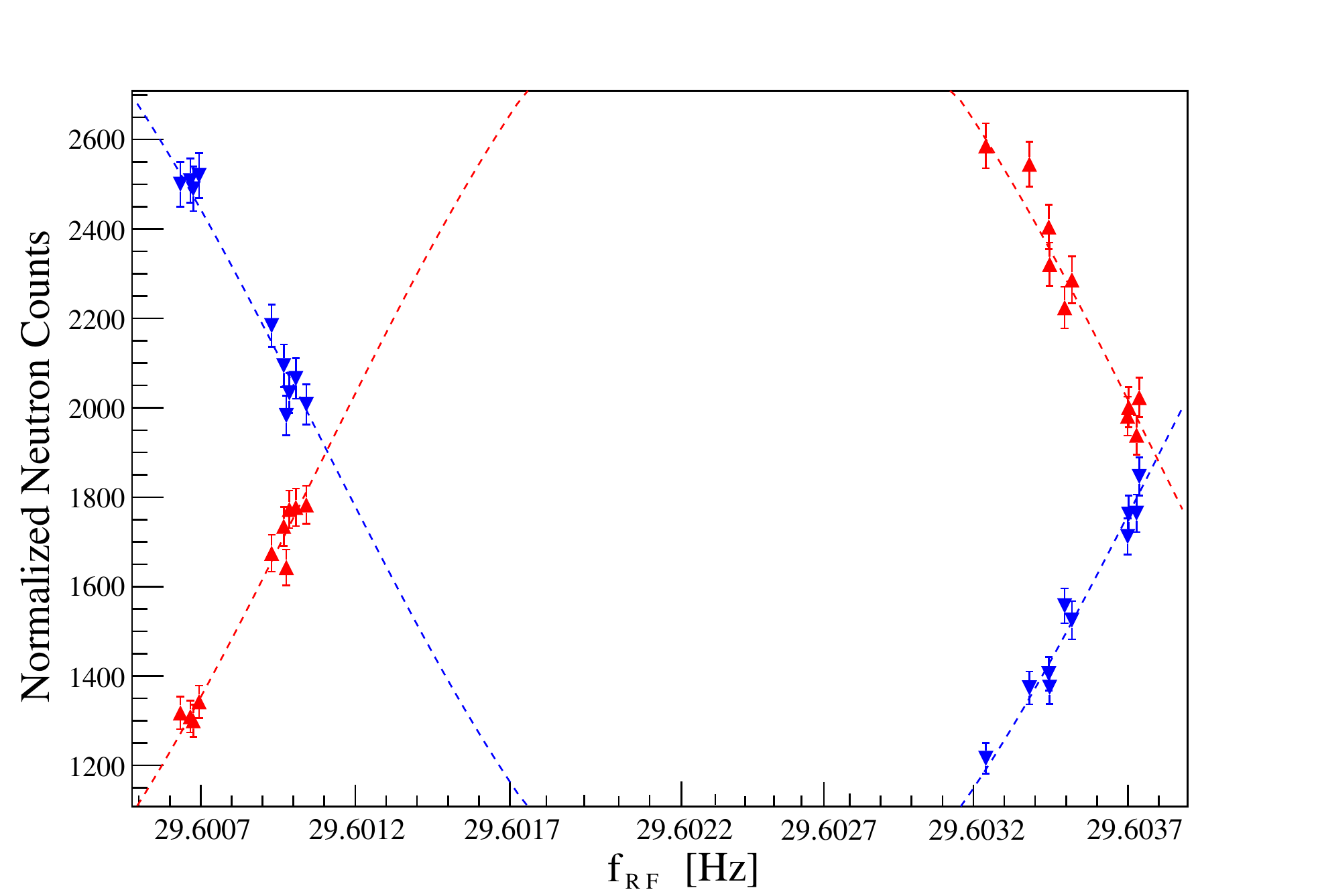}
    \caption{First Ramsey pattern taken with the USSA below the nEDM spectrometer at PSI. It shows the number of detected neutrons of each spin state as a function of $\mathrm{f_{RF}}$. Up triangles are spin up UCN counts and down triangles are spin down UCN counts. Dashed lines are the fitted cosine functions used to extract $N_{\mathrm{tot}}$ and $\alpha$.}
    \label{fig:USSA_Ramsey_pattern}
  \end{figure}
\end{center}

The USSA performance was then compared to that of the former sequential analyser \cite{Baker_2014} (see Fig.\;\ref{fig:spin_analysing_system}) for typical nEDM measurements. The magnetized foil used in the sequential analyser test was made from the same batch as the USSA foils, to eliminate (to first order) foil performance from the comparison. The coating of the sequential analyser is also made of NiMo (with a 85/15, Ni/Mo weight ratio), as for the USSA. The precession chamber was filled for 35\,s and polarised UCN were stored for 180\,s. A Ramsey type measurement \cite{Ramsey_1950} was then performed in order to measure the neutron Larmor frequency $\nu_L$: the UCN spin was initially aligned with the main magnetic field $B_0$. For a 2\,s period, a $\pi/2$ RF pulse, with a frequency $\mathrm{f_{RF}}$, in principle equal to $\nu_L$, was applied. As a result, the neutron spin was flipped into the plane orthogonal to the main magnetic field. Neutrons were then precessing during 180\,s before applying a second $\pi/2$ pulse at the same frequency as the first one for 2\,s. Therefore, the initial UCN spin was flipped after this sequence. The actual Larmor frequency was determined by slightly changing $\mathrm{f_{RF}}$ and by counting the spin up and the spin down UCN, in order to scan the central fringe of the so-called Ramsey pattern, as shown in Fig.\,\ref{fig:USSA_Ramsey_pattern}. The Ramsey central fringe was then fitted using a cosine function to recover the average number of detected UCN $N_{\mathrm{tot}}$ and the fringe contrast $\alpha$. These two parameters determine the figure of merit of the experiment, following Eq.\;\eqref{eq:dn_statistic_error}.

 The fitted values of $N_{\mathrm{tot}}$ and $\alpha$ obtained with the USSA and with the sequential analyser are summarised in Table \ref{tab:EDM_comparison}. With the USSA, both the contrast and the number of detected UCN were larger than with the sequential analyser. The improvement on the nEDM sensitivity was estimated through the variable $\alpha\sqrt{N_{\mathrm{tot}}}$ according to Eq.\,\ref{eq:dn_statistic_error}. It amounts to $(18.2\pm6.1)\%$ with a major contribution coming from the UCN statistics improvement of $(23.9\pm1.0)\%$. This comparison between these systems is realistic as it gives the net increase in the sensitivity, but is idealised, in the sense that a detailed accounting for gap losses in the EDM cell and transport to the analysers was not implemented.
\begin{table}[h]
  \begin{center}
  \caption{Average contrast, $\alpha$, and number of detected UCN, $N_{\mathrm{tot}}$, measured with the USSA and with the sequential analyser for nEDM runs. $N_{\mathrm{mon}}$ is the number of detected UCN on the West-1 beam line used to monitor the number of detected UCN on the nEDM beam line and to correct for UCN source variations.}
    \begin{tabular}{lccc}
     \hline
     \hline 
     & $\alpha$ [$\%$] & $N_{\mathrm{tot}}$ & $N_{\text{mon}} [\times10^6]$\\
     \hline 
     USSA       & $63.4\pm1.8$ & $3791\pm14$ & $1.878\pm0.005$\\
     Sequential & $59.7\pm2.2$ & $2692\pm15$ & $1.651\pm0.016$\\
     Ratio & $1.062\pm0.049$ & $1.239\pm0.010$  &\\
     \hline
     \hline 
    \end{tabular}
  \end{center}
\label{tab:EDM_comparison}
\end{table}

%%_________________________________________________________________________________________________%%
%% Conclusions %%

\section{Conclusions}

In order to improve the statistical sensitivity of the nEDM measurement and reduce possible spin dependent systematic effects, a new simultaneous spin analyser was built, tested and implemented. The goals were threefold: symmetrically treat both UCN spin components, suppress depolarisations and increase the number of detected neutrons.

We have demonstrated that the two separate arms of the USSA treat the two UCN spin states symmetrically. The spin-flipper efficiencies were similar within the 1\,$\%$ level and within the 4\,$\%$ level for the analysing powers of the foils. In addition, the UCN transmission of these arms and the UCN detection efficiency of the two detectors were also similar at the percent level.  Possible cross-talk between the two arms was excluded at the percent level. This last result confirmed preliminary magnetic field measurements. The sole asymmetry between the two arms arose from the non-ideal spin-flipper efficiency of about 97$\%$, that induced a small decrease of the number of detected UCN in the active arm with respect to the opposite arm. Therefore, the shape of the magnetic fields and the RF field strength and profile will be further investigated in order to avoid possible non-adiabatic regions and to increase the spin-flipper efficiencies up to 100$\%$.

Under nEDM running conditions, the increase in the counting statistics of $(23.9 \pm 1.0)\%$ , and the improvement of $(6.2 \pm 4.9)\%$ of the Ramsey fringe contrast lead to an $(18.2 \pm 6.1)\%$ improvement in the nEDM sensitivity over the previously used sequential analyser. Part of the improvement on the counting statistics comes from the reduction of the storage above the analysing system, suppressing losses in the rest of the apparatus. The statistical gain using the USSA will be further investigated during the next nEDM data taking since it is now part of the installed apparatus. It is planned to further improve the UCN transmission in the near future using a coating with a higher Fermi potential than NiMo (220\,neV). For instance for UCN with kinetic energies between $250-300$\,neV the USSA transmission (measured to be 80$\%$) could be increased by the use of $^{58}$NiMo, which is known to have a Fermi potential close to 300\,neV.

\section*{Acknowledgements}

We acknowledge T. Brenner for his warm welcome and his support during the measurements at ILL. The USSA construction would not have been achieved without D.~Gou\-pil\-li\`e\-re, design engineer at LPC Caen, and without the LPC Caen technical staff who carefully machined the pie\-ces. We acknowledge A. Lemari\'e (GANIL) for his assistance during the measurement of the USSA magnetic field on the GANIL measurement bench. We acknowledge M. Horisberger who performed part of the USSA NiMo coating. We are indebted to M. Meyer for preparing the USSA installation at PSI. We also acknowledge the FASTER team for their support. The LPC Caen and the LPSC Grenoble acknowledge the support of the French Agence Nationale de la Recherche (ANR) under reference ANR-09--BLAN-0046. Polish partners acknowledge The Na\-tio\-nal Science Centre, Poland, for the grant No. UMO-2012/\\04/M/ST2/00556. This research was supported in part by PL-Grid infrastructure. The Paul Scherrer Institute and ETH Z\"urich gratefully acknowledge the support of the Swiss National Science Fundation under Grant Nos. 200021\_126562, 200020\_144473 , 200021\_138211, 200020\\\_149211 and CRSII2\_144257.

\bibliography{references}
\bibliographystyle{unsrt}

\end{document}